# PTST: A polar topological structure toolkit and database

Guanshihan Du[1,#], Yuanyuan Yao[2,#], Linming Zhou[1], Yuhui Huang[1, 3], Mohit Tanwani[4], He Tian[1,5,6, 7], Yu Chen[2], Kaishi Song[2], Juan Li[8], Yunjun Gao[2], Sujit Das[4], Yongjun Wu[1,3,6,9,*], Lu Chen[2,*], Zijian Hong[1, 3,5,7,*]

[1] State Key Laboratory of Silicon and Advanced Semiconductor Materials, School of Materials Science and Engineering, Zhejiang University, Hangzhou, Zhejiang 310058, China

[2] College of Computer Science, Zhejiang University, Hangzhou, Zhejiang 310058, China

[3] Zhejiang Key Laboratory of Advanced Solid State Energy Storage Technology and Applications, Taizhou Institute of Zhejiang University, Taizhou, Zhejiang 318000, China

[4] Materials Research Centre, Indian Institute of Science, Bangalore-560012, India

[5] Center of Electron Microscopy, School of Materials Science and Engineering, Zhejiang University, Hangzhou 310027, China

[6] Institute of Fundamental and Transdisciplinary Research, Zhejiang University, Hangzhou 310058, China

[7] Pico Electron Microscopy Center, Hainan University, Haikou 570228, China

[8] College of Materials Science and Engineering, Zhejiang University of Technology, Hangzhou 310014, China.

[9] School of Engineering, Hangzhou City University, Hangzhou, Zhejiang 310015, China

[#] Equal Contributions

[*] Corresponding authors:

Y. W. (yongjunwu@zju.edu.cn); L.C. (luchen@zju.edu.cn); Z. H. (hongzijian100@zju.edu.cn)

**Abstract**

Ferroelectric oxide superlattice with complex topological structures such as vortices, skyrmions, and flux-closure domains have garnered significant attention due to their fascinating properties and potential applications. However, progress in this field is often impeded by challenges such as limited data-sharing mechanisms, redundant data generation efforts, high barriers between simulations and experiments, and the underutilization of existing datasets. To address these challenges, we have created the "Polar Topological Structure Toolbox and Database" (PTST). This community-driven repository compiles both standard datasets from high-throughput phase-field simulations and user-submitted nonstandard datasets. The PTST utilizes a Global–Local Transformer (GL-Transformer) to classify polarization states by dividing each sample into spatial sub-blocks and extracting hierarchical features, resulting in ten distinct topological categories. Through the PTST web interface, users can easily retrieve polarization data based on specific parameters or by matching experimental images. Additionally, a Binary Phase Diagram Generator allows users to create strain and electric field phase diagrams within seconds. By providing ready-to-use configurations and integrated machine-learning workflows, PTST significantly reduces computational load, streamlines reproducible research, and promotes deeper insights into ferroelectric topological transitions.

**Introduction**

The engineering of complex ferroelectric topological structures, such as polar vortices [1-8], flux-closure domains [9-11], spirals [12], skyrmions [13-16], and merons [17], has become a prominent research area across various disciplines, including condensed matter physics and materials science. These structures have greatly enhanced our fundamental understanding of ferroelectric materials and show significant potential for applications in next-generation electronic devices due to their intriguing physical phenomena and properties [18-25]. A well-known example of this is the $PbTiO_3/SrTiO_3$ (PTO/STO) superlattice system, which has been the host for many of the topological phases observed over the past few decades. As new ferroelectric topological structures continue to be discovered, we are entering a critical stage where key challenges related to data sharing have emerged, hindering further developments in this field.

The creation of large sets of comprehensive and open data is a crucial component of the materials genome initiative, which could facilitate the fast materials design when combined with the state-of-the-art artificial intelligence tools. The key challenges in the data acquisition and sharing of polar topological structures includes: 1. Redundant generation of identical data, which occurs when different researchers replicate the same parameter sweeps and growth procedures. 2. The absence of an effective data-sharing infrastructure, which limits the accessibility and reproducibility of essential datasets. 3. Insufficient data-driven insights, as many simulation outputs remain underutilized after addressing immediate research questions. These issues underscore the urgent need for a database to aggregate and distribute high-quality data on polar domains in ferroelectric superlattices.

To address these challenges, we developed an open data platform called the "Polar Topological Structure Toolkit and Database" (PTST), which integrates standard and nonstandard datasets to encompass a wide range of ferroelectric polarization configurations in the ferroelectric superlattice system. The standard database is built via high-throughput phase-field simulations that systematically vary key parameters (e.g., superlattice thickness, substrate strain, and electric field), minimizing repetitive data generation. Meanwhile, the nonstandard database collects user-submitted computational and experimental configurations, further broadening the research scope. PTST employs a Global–Local Transformer (GL-Transformer)—a neural network architecture that is widely employed in both NLP[29] and medical image analysis[30]—to classify representative ferroelectric configurations. This classification framework powers features on the "PTST" website, including two-dimensional phase diagram generation, direct downloads of polarization data, and an image-based search module that could match experimental images to simulation data. We hope to spur further interest in the development of high-quality database for ferroelectric topological phases.

**Main**

The construction and application of the PTST are illustrated in **Figure 1**. As depicted in Figure 1(a), PTST collects two types of data: 1. A standard dataset generated through high-throughput phase-field simulations, which systematically vary four key parameters, including superlattice layer thickness, applied voltage, and substrate strain along the *X*- and *Y*-directions (for more details, refer to the **Methods** and **Supplementary Information**). 2. A nonstandard dataset that comprises user-uploaded spatial polarization data via the official PTST website. The standard dataset includes 2,541 polarization configurations stored in .in format (updated regularly with new data). These are then compressed into .npz format to ensure high-fidelity storage and easy retrieval. The nonstandard dataset features both computed and experimentally measured polarization configurations and morphologies, along with optional parameters and descriptive information. Once vetted by database administrators, these user-generated data can significantly enrich PTST by expanding its data space.

Figure 1(b) presents the PTST website interface, which allows users to customize parameters such as strain, electric field components, and system size to quickly generate initial polarization structures. Users can also upload their non-standard datasets. Both types of data undergo feature extraction through a Global–Local Transformer (GL-Transformer) that is pre-trained using self-supervised learning (for more details, see **Methods** and **Figure. S1**). The GL-Transformer divides each sample into four blocks along the *XY* plane and further splits them into upper and lower layers along the Z-axis, resulting in eight total sub-blocks. Each sub-block is processed by a BlockTransformer module, which produces a 128-dimensional embedding. These embeddings are then combined into a single global representation. The training curve of the GL-Transformer model is shown in **Figure. S2**. This network architecture offers three primary advantages: (1) it models global context through attention mechanisms, capturing spatial dependencies and interactions among the three polarization components; (2) it dynamically focuses on important regions within each sub-block, ensuring that noisy or sparse data do not overshadow critical features; and (3) it enables multiscale feature extraction through a stacked-encoder structure, allowing the identification of hierarchical patterns from local domains to the entire superlattice film.

A hierarchical clustering algorithm classifies each global embedding into one of ten polar structure categories, which include vortices, skyrmions, and various other domain structures. This classification system creates a robust taxonomy for data retrieval and analysis. The PTST platform supports a variety of key applications (see Figure 1c), including structure construction, topological domain analysis, tracking polarization evolution, integrating machine learning, and enabling data sharing. By utilizing these resources, researchers can streamline complex workflows in ferroelectric materials research, from generating and categorizing polarization data to conducting advanced computational analyses and predictive modeling.

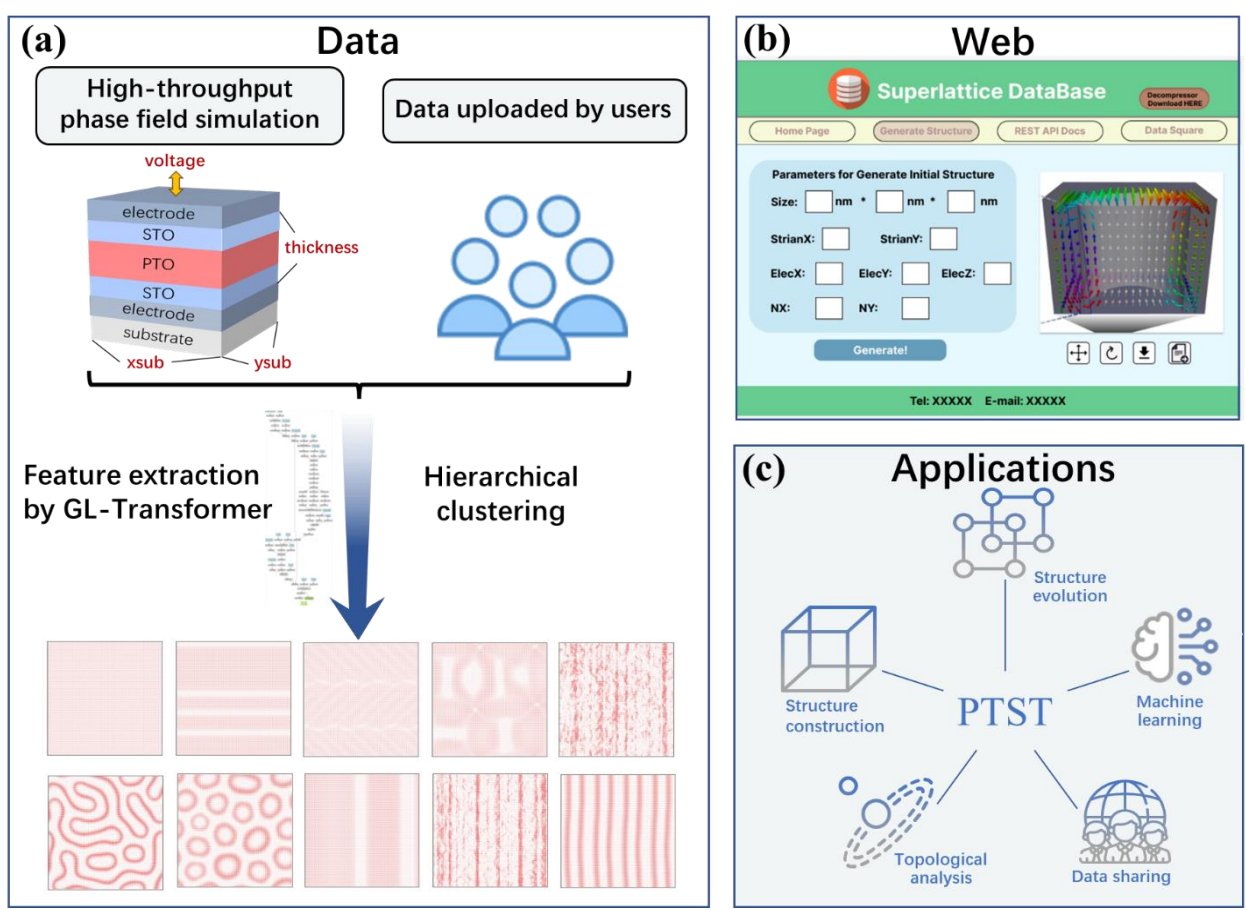

**Figure 1.** Schematic representation of the PTST construction and application process. (a) The PTST database consists of a standard dataset generated via high-throughput phase-field simulations by varying key parameters such as superlattice layer thickness, applied voltage, and substrate strain, and a nonstandard dataset enriched by user-contributed polarization data. Both datasets undergo feature extraction using a GL-Transformer network, followed by hierarchical clustering to categorize polarization configurations. (b) The "PTST" website interface allows users to generate initial structures by customizing parameters such as strain, electric field components, and system size. (c) The PTST platform supports a range of applications including structure

construction, topological analysis, structure evolution, machine learning, and data sharing, providing comprehensive resources for ferroelectric material research.

Building on the deep-learning feature extraction previously described, each data sample's 128-dimensional embedding undergoes hierarchical clustering for classification, as illustrated in **Figure 2**. An agglomerative clustering algorithm employing Ward linkage is utilized, which optimizes a variance-minimizing approach by successively merging clusters based on reductions in their within-group sums of squares. This process is entirely unsupervised, requiring no predefined labels and allowing the data itself to dictate the formation of clusters. Ward's linkage is particularly advantageous for the polarization data that may present significant heterogeneity, which tends to yield relatively compact and homogeneous clusters. To demonstrate how different numbers of clusters impact the data distribution, a principal component analysis (PCA) is conducted to reduce high-dimensional embeddings into two principal components (see Figure 2a). By visualizing the data points in this two-dimensional space, it becomes easier to identify cluster boundaries and observe how varying the target cluster count could affect group separations. Different colors and shapes represent clusters at each stage (original data, 2, 4, 6, 8, and 10 clusters), which helps in interpreting how gradually increasing the number of clusters reveals more subtle patterns within the data. After consolidating the embeddings, a hierarchical process generates a dendrogram (Figure 2b), which is used to determine the final ten clusters by establishing an appropriate cutting plane. The dendrogram offers a nested perspective of how the data form subgroups, illustrating which embeddings merge at each level without relying on any pre-labeled categories. This clustering framework ultimately results in ten categories of polarization patterns that align with the domain structures and topological features identified in the earlier analysis. As confirmed by the PCA projection and the dendrogram, the resulting clusters exhibit high internal consistency and clear distinctions between clusters. This underscores the effectiveness of GL-Transformer embeddings in capturing essential structural variations in configurations such as vortices, skyrmions, and other ferroelectric domains.

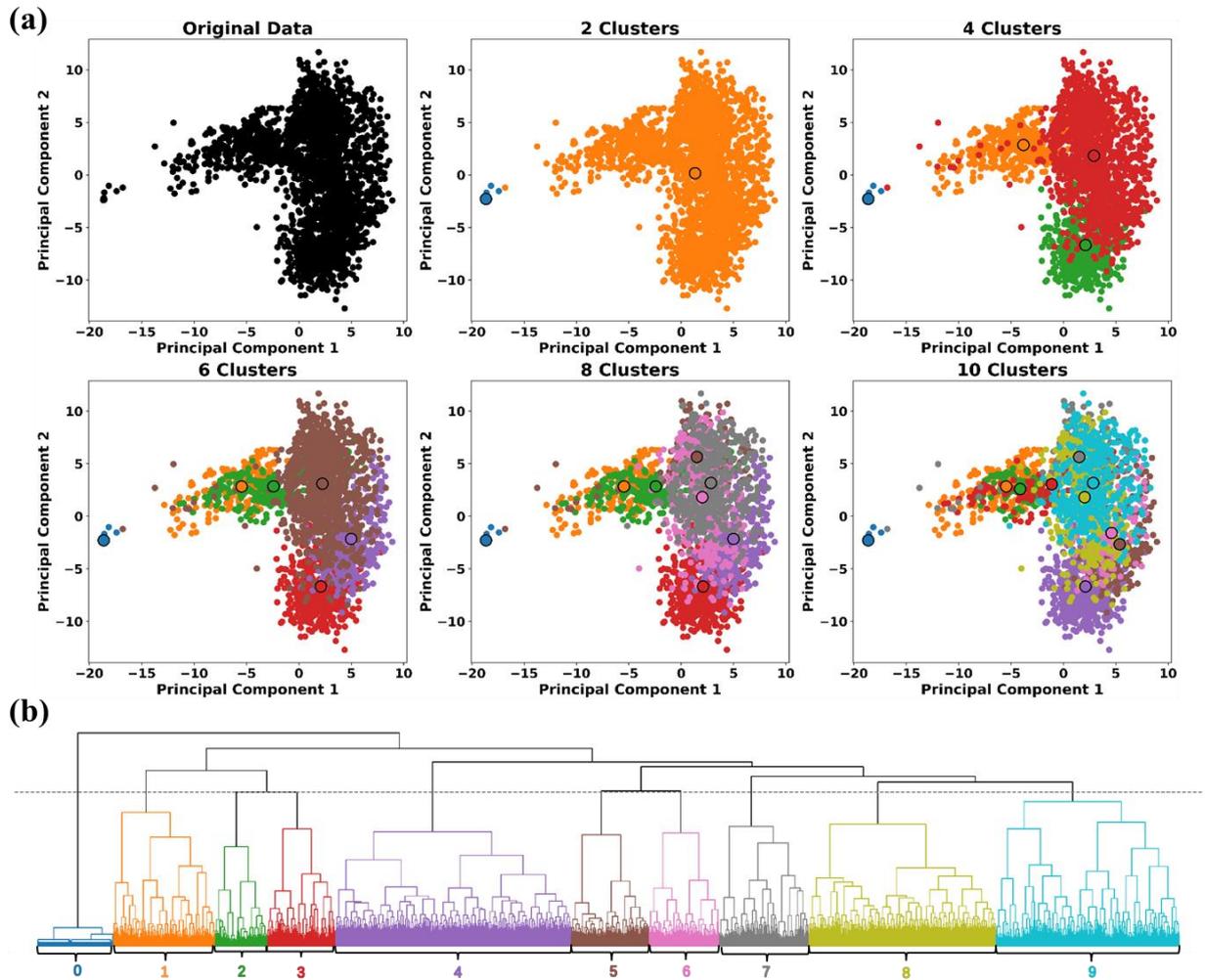

**Figure 2.** Hierarchical clustering and visualization of polarization data embeddings. (a) Principal component analysis (PCA) reduces 128-dimensional embeddings into two dimensions, illustrating data separation at varying cluster counts (2, 4, 6, 8, and 10 clusters). (b) Dendrogram generated via agglomerative clustering with Ward linkage reveals hierarchical relationships, enabling the selection of ten final clusters that represent distinct polarization patterns.

We then proceed to investigate the details of the ten distinct polarization categories identified through hierarchical clustering within the PTST standard database, as illustrated in **Figure 3**. Figures 3(a-j) display representative 3D renderings of these ten cluster types, showcasing not only three simple single-domain states with varying orientations but also more complex configurations, such as a/c domains, a1/a2 domains, labyrinth patterns, skyrmions, vortices, and mixed structures, corresponding in-plane and out-of-plane polarization distribution maps for these polarization structures are provided in **Figure S3**, respectively. Both theoretical and experimental observations support the existence of these different polarization states. Notably, our model demonstrates the ability to handle large-scale 3D mixtures of complex polar topologies. Figure 3(k) quantifies the overall distribution of samples assigned to each cluster, labeled 0–9, corresponding to the rendered structures above. It is evident that the data across the ten categories are almost evenly distributed, although the single-domain state has a slightly higher population compared to the complex polar topologies. As a result, users of the PTST can easily locate and compare different polarization states, enabling both targeted investigations into specific domain phenomena and broader surveys of topological diversity.

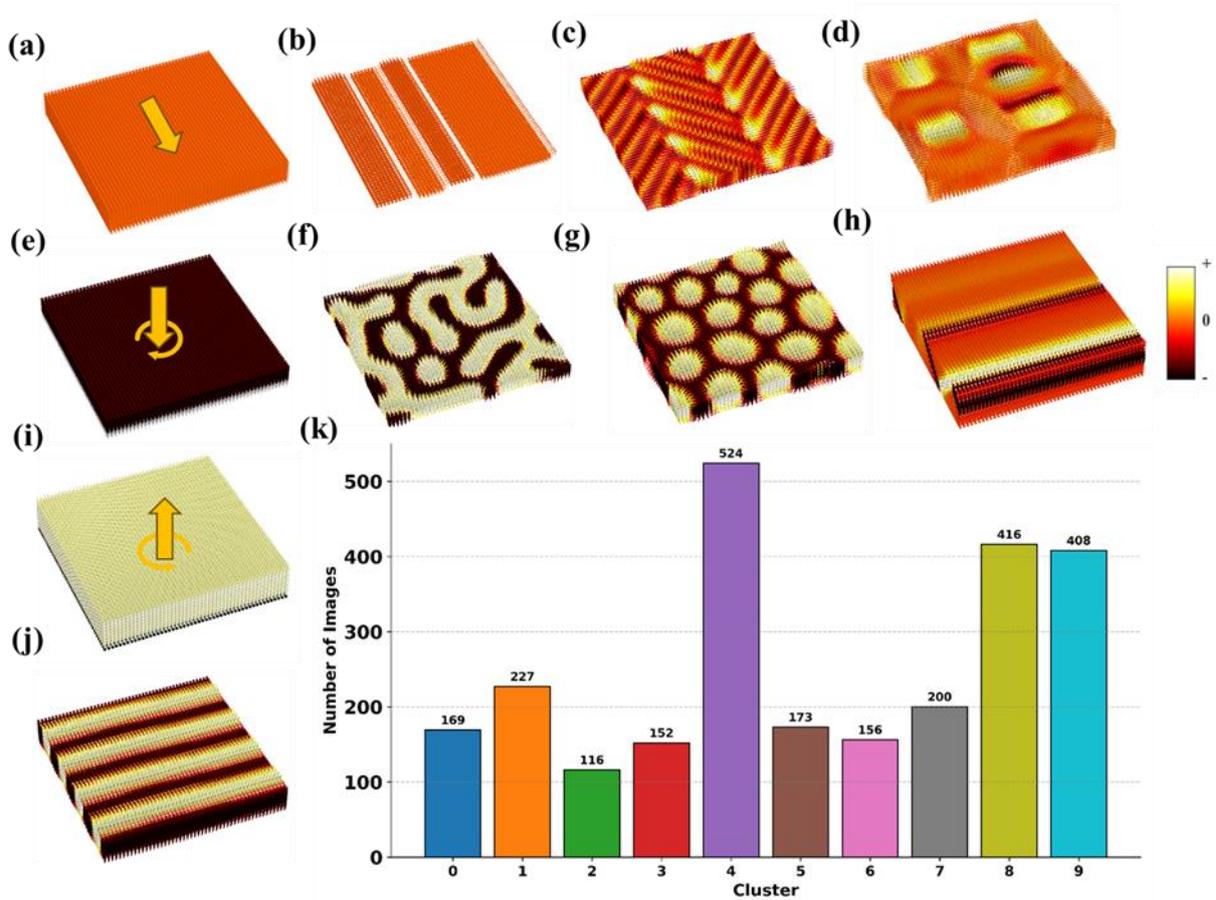

**Figure 3.** Visualization and distribution of polarization structures in the PTST standard database. (a–j) 3D renderings of ten clustered polarization structures: (a, e, i) single domains with different orientations, (b) a1+/a1- domain, (c) a1/a2 domain, (d) mixed domain, (f) labyrinth, (g) skyrmion, (h) a/c domain, and (j) vortex. (k) Distribution of images among the ten clusters, with categories 0–9 corresponding to structures shown in (a–j).

An online open-source platform has been developed to streamline the generation and retrieval of polarization configurations from PTST, as illustrated in **Figure 4**. Users can access polarization data through two methods: (1) Precise search: This method allows users to input specific parameters, such as $x_{sub}$, $y_{sub}$, voltage, and superlattice thickness. By doing so, they can directly obtain the corresponding polarization data, cross-sectional images, and 3D polarization maps from the standard database. (2) Fuzzy search: In this approach, users provide a target topological structure along with approximate parameter ranges. The system first identifies the relevant structural category and then performs a parameter-based match. These capabilities significantly reduce the necessity for lengthy phase-field simulations to generate initial domain states from scratch, thereby accelerating research workflows and materials screening. Figure 4(a-b) illustrates the rapid search for two classical topological structures that have been widely observed: the polar vortex and the polar skyrmion, using two different sets of specified parameters. The resulting polar patterns serve as inputs for phase-field simulations, which follow a standard procedure to relax into equilibrium. The energy evolutions for both the vortex (Figure 4c) and skyrmion (Figure 4d) configurations are shown under two different initialization schemes: one starting from random noise and the other from the polarization data extracted using the PTST database. The red curves represent simulations that begin with PTST-generated data, demonstrating a quick relaxation to a stable energy state, typically within 500 timesteps. In contrast, the blue curves, which are initialized from random noise under the same conditions, take over 5,000 timesteps to reach the same equilibrium state. This significant difference highlights how PTST can dramatically reduce computational costs, allowing researchers to focus on scientific exploration rather than repetitive data generation. Additionally, the platform's flexible design enhances reproducibility in the field by allowing consistent reuse of initial structures across parallel or comparative studies. Furthermore, this database is user-friendly for beginners who may not be well-versed in the thermodynamic conditions necessary to stabilize these different phases.

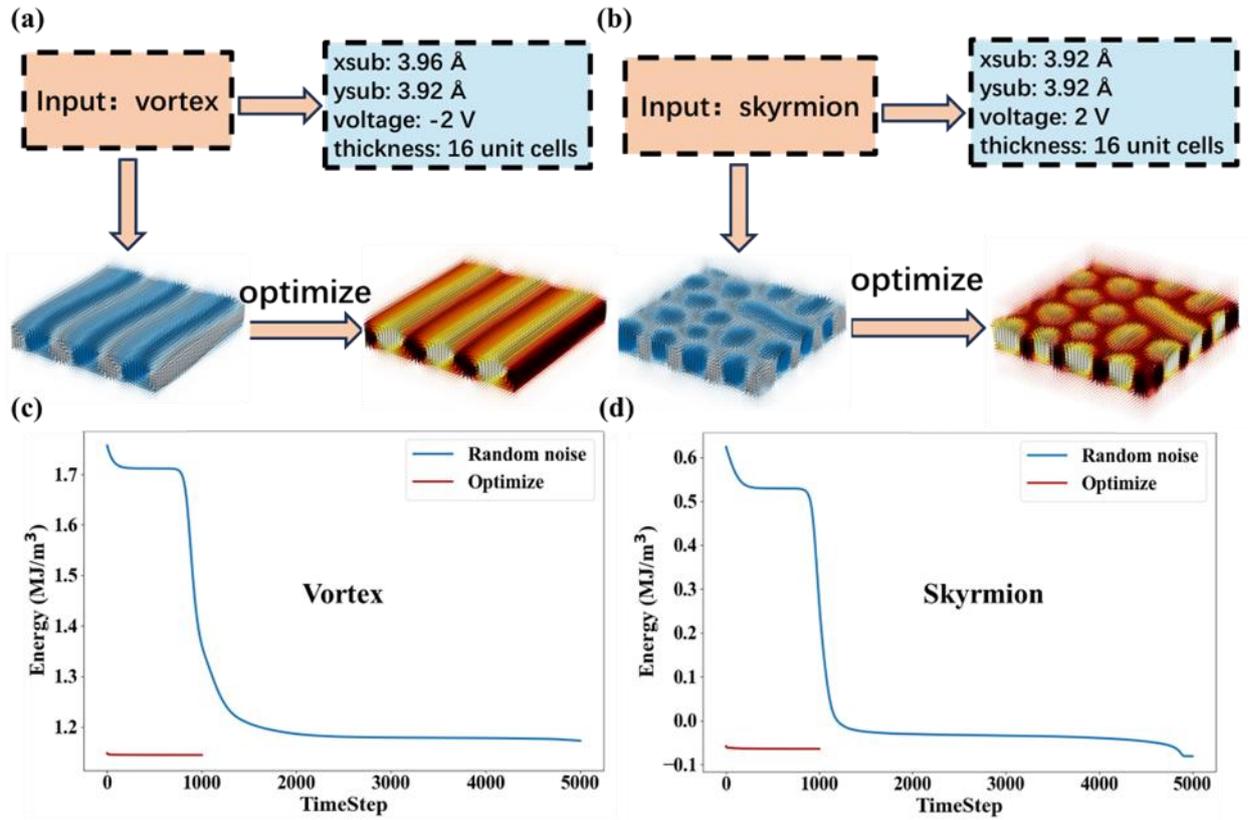

**Figure 4. Demonstration of PTST's Polarization Data Generator.** (a, b) Users can select a desired topological structure, such as vortex or skyrmion, and the system generates the corresponding spatial polarization data along with related parameters (xsub, ysub, voltage, and thickness). The generated data can be directly used as input for the user's phase-field simulation program for further optimization. (c, d) Energy evolution during phase-field simulations for the vortex (c) and skyrmion (d) configurations. The red curves represent energy relaxation using the PTST-generated polarization data, while the blue curves show energy relaxation starting from random noise under the same conditions.

We have developed an image-based search module within the Polarization Data Generator, in addition to the existing parameter-based retrieval system, as illustrated in **Figure 5**. This module utilizes a deep learning model to match experimental polarization images—such as those obtained from Transmission Electron Microscopy (TEM) or Piezoresponse Force Microscopy (PFM)—with a standard database that we have created. As shown in Figure 5a, users can upload cross-sectional and planar view TEM images (an example for the polar vortex phase is taken from Ajay et al. [4]). The system identifies the closest match to a vortex polarization configuration, displaying the corresponding XY and XZ cross-sections, 3D polarization distribution, and the parameters associated with generating this structure. Similarly, in Figure 5b, when users upload a planar view TEM image alongside a PFM image, the system matches them to a skyrmion polarization configuration and provides the relevant simulated images and parameters. In this workflow, each uploaded image undergoes contrast-optimized preprocessing to ensure that the TEM and PFM signals are adequately normalized. A feature-extraction neural network encodes the image into a high-dimensional vector, which is then compared using cosine similarity to precomputed feature libraries of simulated thin-film sections from the database. Details about the underlying machine learning model and its implementation can be found in the **Methods** section. Once the best match is identified, the platform provides users with cross-sectional snapshots (XY and XZ views), a three-dimensional polarization map, associated structural parameters (such as layer thickness, substrate strain, and voltage), and options to download the matching polarization data directly. These data allow researchers to immediately apply the discovered configuration in further phase-field simulations, eliminating the need for extensive trial and error to reconstruct similar states. By offering cross-validation against experimental images and rapid access to closely matched simulated structures, this module significantly accelerates the research cycle for ferroelectric heterostructures.

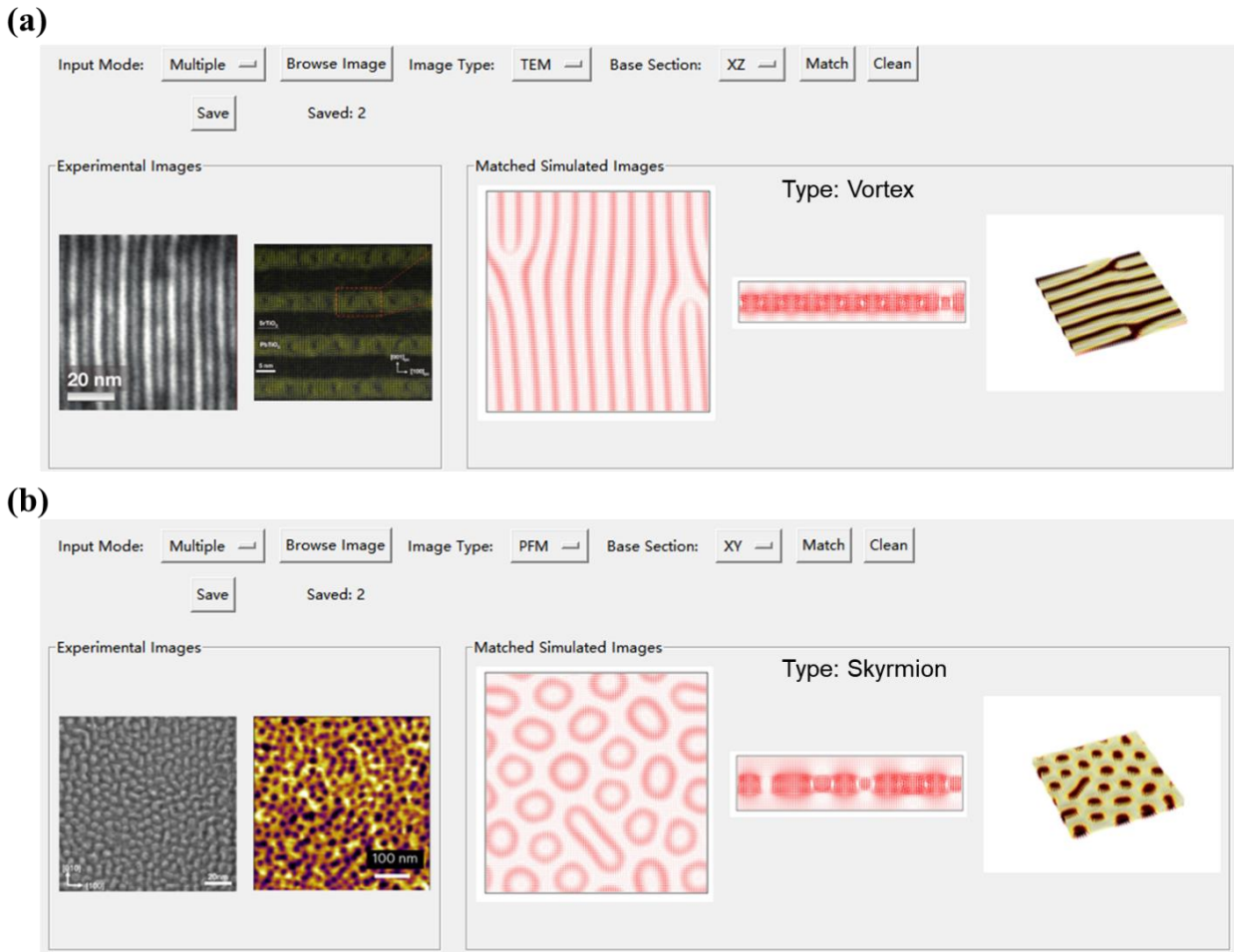

**Figure 5. Demonstration of PTST's Image-Based Polarization Matching Module.** (a) The user uploads TEM images of the thin film's XZ and XY cross-sections, and the system identifies the closest match as a vortex polarization configuration, displaying its corresponding XY and XZ cross-sections, 3D polarization distribution, and associated parameters. (b) The user uploads an XY-section TEM image and a PFM image, and the system matches it to a skyrmion polarization configuration, providing the corresponding simulated images and parameters.

We have further developed a toolkit called the Binary Phase Diagram Generator for predicting phase diagrams using data from the PTST database, as shown in **Figure 6**. This tool allows users to select two variables to represent the axes of the diagram (for example, $y_{sub}$ on the *x*-axis and voltage on the *y*-axis), while other parameters (such as superlattice thickness or strain value in a different direction) can be fixed based on the experimental or theoretical needs. Once these selections are made, the tool generates a preliminary two-dimensional phase diagram that color-codes different polarization states, with each color corresponding to one of the ten identified cluster labels. By mapping the transitions between various domain configurations, this toolkit provides valuable insights into how subtle changes in external or structural variables can affect the resulting polarization landscapes. Researchers can effectively identify critical boundaries that separate distinct phases and determine the optimal conditions for stabilizing specific topological domains. Additionally, the diagrams produced can be easily updated or refined when new data is added to the PTST database, making it easy to timely incorporate emerging material systems. This level of flexibility and real-time responsiveness significantly speeds up the hypothesis testing process, particularly for studies that require rapid exploration across a wide range of parameters to discover novel ferroelectric phenomena or to optimize design parameters for future materials design. This user-friendly tool can be utilized by experimentalists worldwide, even those without a background in phase-field simulations.

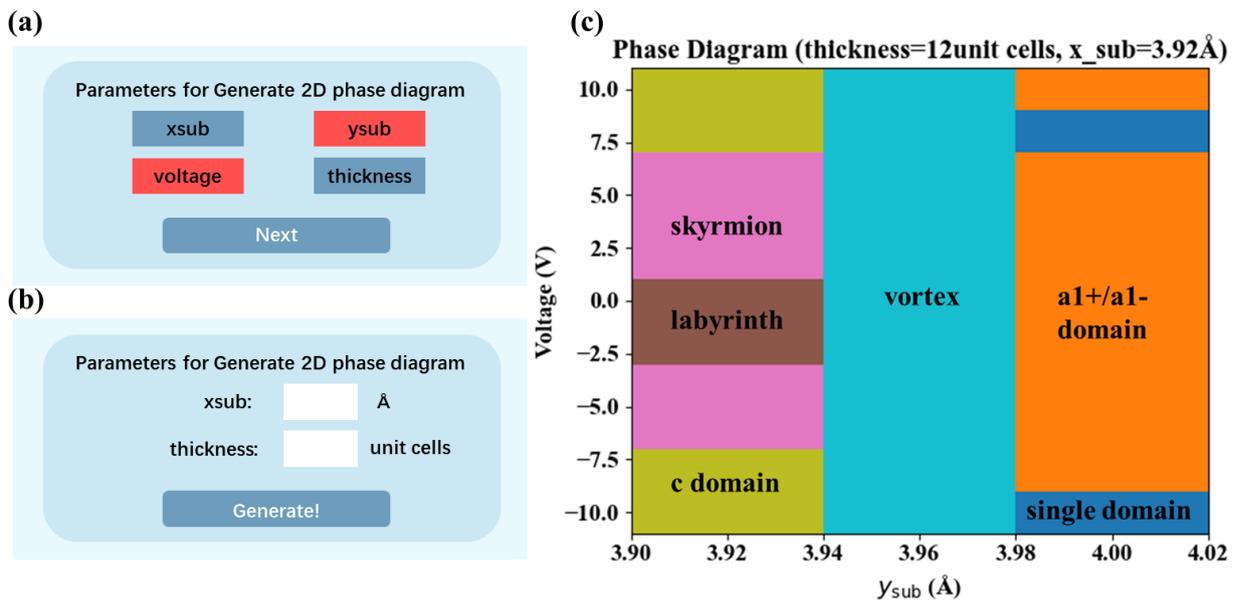

**Figure 6.** Demonstration of PTST's Binary Phase Diagram Generator. (a) Users select two variables for the phase diagram's axes (e.g., x-axis: ysub, y-axis: voltage). (b) The remaining variables (e.g., thickness: 12 unit cells, xsub: 3.92 Å) are set. (c) The generated binary phase diagram shows distinct phase regions represented by different colors, each corresponding to a specific domain structure, as shown in Figure 3(a-j).

In summary, PTST offers a comprehensive repository of ferroelectric spatial polarization configurations along with their associated metadata. This repository is organized using a dual-structured design that includes both standard and nonstandard datasets. The standard dataset is generated through high-throughput phase-field simulations, while the nonstandard datasets are continuously expanded through user contributions. The Global–Local Transformer (GL-Transformer) architecture, combined with agglomerative clustering, effectively captures the complexity of these polarization states and identifies ten distinct categories, including vortex and skyrmion structures, as well as various other complex domain structures discovered in the PbTiO$_3$/SrTiO$_3$ superlattice system. Principal component analysis and dendrogram visualizations further validate the consistency and separability of these clusters. In addition to static categorization, the PTST platform provides both parameter-based and image-based retrieval of spatial polarization data, allowing rapid construction of customized domain configurations. The ready-to-use data significantly accelerate phase-field simulations and minimize unnecessary computational efforts. Notably, the platform's image-based search module enables researchers to upload experimental TEM or PFM images, allowing them to quickly find the closest matching polarization states. This feature could effectively bridge the gap between simulation and experimentation. Additionally, the Binary Phase Diagram Generator module identifies critical boundaries between different polarization regimes and determines optimal parameters for stabilizing desired topological states. By integrating both expert-curated and user-generated data, along with an intuitive interface, PTST remains adaptable to various research needs, supporting studies ranging from fundamental ferroelectric physics, materials-by-design to device-oriented explorations. Looking ahead, we plan to further expand PTST by incorporating more physical variables, such as temperature, stress, and optical pumping. Moreover, machine learning tools will be further developed to enhance the predictive power of the toolkits. Streamlined compatibility with advanced simulation software and automated workflows will enhance the efficiency of updating, analyzing, and visualizing evolving datasets. Ultimately, PTST aims to be a continually growing, community-driven resource that drives discoveries in ferroelectric topological materials, encourages collaborative innovation, and paves the way for next-generation theoretical and experimental breakthroughs.

**Acknowledgement**

The financial supports from Natural Science Foundation of Zhejiang Province (No. LR25E020003, ZH; No. LD24E020003, YW), the National Natural Science Foundation of China (No. 92166104, No. 92463306, ZH), and the Joint Funds of the National Natural Science Foundation of China (No. U21A2067, YW) and are acknowledged. ZH is supported by the Fundamental Research Funds for the Central Universities (2023QZJH13). The simulation results in this work were obtained using the Mu-PRO software package (https://muprosoftware.com).

**Methods**

*High-Throughput Phase-Field Simulations*

The spontaneous polarization vector ($\vec{P}$, i = 1, 3) was adopted as the order parameter, and its

temporal evolution was described by the time-dependent Ginzburg–Landau (TDGL) equation [31-33]:

$$\frac{d\vec{P}}{dt} = -L\frac{\delta F(\vec{P})}{\delta \vec{P}}$$

where $\vec{P}$ represents the spontaneous polarization vector, $t$ is the evolution time step, and $L$ denotes the kinetic coefficient related to domain wall mobility. The total free energy $F(\vec{P})$ includes contributions from Landau theory, elastic effects, electric fields, and gradient energies, integrated over the film volume $V$:

$$F(\vec{P}) = \int (f_{elas} + f_{elec} + f_{grad} + f_{land})\, dV$$

Further details about the energy formulations and simulation constants can be found in earlier studies [7, 14, 31, 34]. In this research, periodic boundary conditions were implemented in the in-plane dimensions, while a superposition method was applied in the out-of-plane ($z$) direction []. More specific details on the energy formulations and simulation constants are available in earlier studies [7,14,31,34]. We considered a trilayer structure of $(STO)_n/(PTO)_n/(STO)_n$, where $n$ varies from 4 to 24 in increments of 2. The simulation system was discretized into a 3-D grid of 100×100×Nz, where Nz is determined by the sum of the substrate (30 grid points), the superlattice film ($3n$ grid points), and the air above the superlattice (30 grid points) along the out-of-plane axis. An iterative perturbation method is used to consider the elastic anisotropy for the PTO and STO layers [35]. The normalized time step is set as 0.01 in this study.

A Python script was developed to systematically vary four key parameters:

1. Periodicity $n$: varies from 4 to 24 in increments of 2.

2. Applied voltage (V): ranges from −10 V to +10 V in step of 2 V.

3. Substrate lattice parameter in $x$-direction ($x_{sub}$) and $y$-direction ($y_{sub}$): spans from 3.80 Å to 4.00 Å in increments of 0.04 Å), with the condition that $x_{sub} \geq y_{sub}$ to avoid redundant simulations.

For each unique combination of these parameters, corresponding phase-field input files were automatically generated. The elastic anisotropy was addressed through an iterative solver to accurately incorporate strain effects under the specified substrate conditions. The 3D polarization distributions obtained from these simulations were initially stored in .in files and subsequently compressed into .npz format to enhance storage efficiency. In total, this workflow produced 2,541 simulation data, covering a wide range of substrate strain states and applied electric biases.

*GL-Transformer Feature Extraction and Hierarchical Clustering*

Each .npz file generated by the high-throughput phase-field simulations is first decompressed into a .in file, restoring the three-dimensional grid coordinates and polarization components (X, Y, Z, Px, Py, Pz). Once decompressed, the region of interest (for example, the PTO layer in a PTO/STO superlattice) is selectively extracted according to thickness and other user-defined parameters.

A self-supervised model (BlockTransformerSSLPE) is then used to learn representative features of the local polarization structure. This model employs sinusoidal positional encoding in three dimensions, implemented via PyTorch, to incorporate spatial information. Specifically, each data row [X, Y, Z, Px, Py, Pz] is augmented with an embedding of its (X, Y, Z) coordinates at multiple frequencies, which is then projected into a Transformer encoder. A self-supervised objective is applied to recover masked entries at both the row level (e.g., position and polarization) and block level (e.g., mean or range of Px, Py, Pz), improving the model's capacity to capture both local and global features.

For feature extraction, a BlockTransformerEvalPE module uses the same positional encoding and Transformer encoder layers but removes the training heads. Each sample's (100×100×Nz) grid is divided into four blocks in the (x,y) plane, then further split at the mid-plane along z, generating eight sub-blocks. Each sub-block's embedding is computed independently and subsequently aggregated via a lightweight AggregatorTransformer, which applies an internal Transformer encoder with attention pooling to produce a final (1×128) embedding for the entire sample.

After generating these embeddings, the ferroelectric configurations are clustered using agglomerative clustering (Ward linkage). A dendrogram is constructed from the resulting linkage, and the cutting plane determines the final number of clusters (in this study, ten). Principal component analysis (PCA) is performed to reduce the high-dimensional embedding space into two dimensions for visualization. All code is written in Python, integrating *PyTorch* for the Transformer-based architecture, *Numpy* and *Scipy* for data manipulation and clustering, and *Matplotlib* for plotting and data visualization.

*Image matching model design*

The image matching and feature extraction module consists of three core components: model pre-training, feature extraction, image matching. Experiment parameter parsing and the graphical user interface (GUI) are implemented to facilitate querying and demonstration. The system achieves a complete workflow from image processing to the presentation of matching

results through the collaboration of these modules. The detailed description of each module is as follows:

- Model Pre-training: We use the pre-trained ResNet18 model and fine-tune it based on the images in the dataset (containing TEM and PFM labels). During the fine-tuning process, the fully connected layers of ResNet18 [36] are replaced with an Identity layer, allowing the model to output intermediate layer feature vectors. These feature vectors will serve as the basis for subsequent image matching.

$$f(x) = Identity(ResNet18(x)),$$

where x is the input image. $ResNet18(x)$ represents the feature extraction process through the ResNet18 model, including multiple convolution layers, residual blocks, and pooling operations. The `Identity` layer at the end retrieves the feature vector from the penultimate layer of the model instead of the traditional classification output.

- Feature Extraction: During the image preprocessing stage, the image contrast is adjusted based on its category (TEM or PFM). The image is first converted to grayscale and then equalized. For TEM images, the colorize function maps them to a white background with a red highlight, whereas for PFM images, a softer red highlight is applied. Then, each image undergoes preprocessing, including resizing, normalization, etc. Afterward, the fine-tuned ResNet18 model is used as the feature extractor to extract the feature vector of the image, which is then $L_2$ normalized.

- Image Matching: For the input experimental image, we calculate the cosine similarity between its feature vector and the feature vectors of each image in the reference feature database. The cosine similarity is calculated using the following formula:

$$consine_{similarity}(v_1, v_2) = \frac{v_1 \cdot v_2}{||v_1|| \cdot ||v_2||},$$

where $v_1$ and $v_2$ are the feature vectors of the experimental image and the reference database image, respectively. The most similar image is returned.

After the model is trained, relevant experimental parameters such as thickness, external electric, and others parameters are extracted from the input filename. These parameters assist in categorizing and managing experiments effectively. Additionally, a graphical user interface (GUI) is developed to allow users to select experimental images, perform matching, and view the results directly within the interface. The system automates the entire workflow, including image loading, preprocessing, feature extraction, similarity calculation, and result presentation.

*Database design*

The material database includes a standard database with high-throughput computational data and a non-standard database for user-uploaded computational and experimental data. The data for both the standard and non-standard databases are stored in the same 'grid_data' table (see Table 1), and distinguished by the 'label' column, where 0 represents standard data and 1 represents non-standard data. The standard database is built from equidistant grid data generated via high-throughput computing, where each row represents a grid point containing key parameters such as the size external electric field components ('elecX', 'elecY', 'elecZ'), and substrate strain ('strainX', 'strainY'). Fields like 'sizeX', 'sizeY', and 'sizeZ' provide precise descriptions for each data point of high-throughput phase-field simulations, ensuring data integrity and consistency. The table design follows a unified processing protocol, optimizing the storage structure to ensure efficient storage and fast retrieval. Additionally, fields like 'name', 'XY_fig', and 'XZ_fig' store file names and image data paths to provide detailed data information during queries.

The non-standard database mainly consists of user-uploaded data, so the table design is more flexible, supporting various data formats. The information of user-uploaded files is stored in the 'data_file' field. To manage non-standard data, a dedicated upload record table is designed, which records the file type, data format, upload date, etc. (see Table 2), enabling flexible management and accommodating different data formats.

We design different index structures for the two database based on various query scenarios, as follows:

- Composite Unique Index: Applied to the combination of (NX, NY, strainX, strainY, elecZ fields in the grid_data table. This ensures that the simulation results are unique for each combination of physical conditions.
- Single-Column Index: Applied to fields like NX, NY, strainX, strainY, in the grid_data table. This improves the query efficiency for these individual fields.
- Foreign Key Index: Applied to the grid_id field in the upload_data table. This accelerates join queries between the upload_data table and the grid_data table.
- Enumeration Field Index: This applies to the 'type' field in the grid_data table and the 'data_type' and 'data_format' fields in the upload_data table. It improves the speed of queries that filter by data type and format.

Standard Database queries efficiently retrieve simulation results by using composite and single-column indexes for parameter-based search, allowing for fast access based on layer thickness, voltage, and strain values. Type-based searches are enhanced by enumeration field indexes, enabling quick filtering by polarization structure types. Image-based search allows users to upload images for automatic matching with simulated data. Fast joins are achieved through foreign key indexing on 'grid_id', improving query performance between polarization and simulation data.

For non-standard databases, flexible tag and fuzzy search capabilities support retrieval with keyword and parameter filters. Enumeration field indexes are used for quick data-type filtering, enabling fast queries based on data format and type. Additionally, category-based search allows filtering by system types such as 'Experiment" or 'Calculation', further enhancing query accuracy.

Table 1. Grid data table design (Table: grid_data)

| Field Name | Data Type | Description | Constraints |
| --- | --- | --- | --- |
| grid_id | BIGINT | Unique identifier for each grid point | Primary Key, Auto Increment |
| name | VARCHAR(255) | The name of file | Not null |
| sizeX | DECIMAL(10, 5) | The size of high-throughput phase-field simulations | Not null |
| sizeY | DECIMAL(10, 5) | The size of high-throughput phase-field simulations | Not null |
| sizeZ | DECIMAL(10, 5) | The size of high-throughput phase-field simulations | Not null |
| NX | DECIMAL(10, 5) | Ferroelectric layer thickness | Not null |
| NY | DECIMAL(10, 5) | Paraelectric layer thickness | Not null |
| strainX | DECIMAL(10, 5) | X-direction substrate strain | Not Null |
| strainY | DECIMAL(10, 5) | Y-direction substrate strain | Not Null |
| elecX | DECIMAL(10, 5) | The size of the external electric field component | Not Null |
| elecY | DECIMAL(10, 5) | The size of the external electric field component | Not Null |

| Field Name | Data Type | Description | Constraints |
| --- | --- | --- | --- |
| elecZ | DECIMAL(10, 5) | The size of the external electric field component | Not Null |
| XY_fig | VARCHAR(255) | XY cross-sectional view of the ferroelectric layer | Not Null |
| XZ_fig | VARCHAR(255) | XZ cross-sectional view of the ferroelectric layer | Not Null |
| data_file | DECIMAL(20, 10) | Data file | Not Null |
| type | INTEGER(4) | Polarization structure type | NOT Null |
| label | INTEGER | Database type | Not Null |
| created_at | TIMESTAMP | Data creation time | Default current time |
| updated_at | TIMESTAMP | Data update time | Default current time |

Table 2. Upload data table design (Table: upload_data)

| Field Name | Data Type | Description | Constraints |
| --- | --- | --- | --- |
| upload_id | BIGINT | Unique identifier | Primary Key |
| user_id | BIGINT | Reference the user table | Foreign Key |
| data_type | ENUM('Experiment', 'Calculation') | Data type | Not Null |
| data_format | ENUM('CSV', 'JSON', 'TXT', 'XML') | Data format | Not Null |
| file_path | VARCHAR(255) | Data file storage path | Not Null |
| state | INTEGER(4) | Upload state | Not Null |
| grid_id | BIGINT | The gird_id | Not Null |
| upload_date | TIMESTAMP | Reference the grid_data | Default current time |

| approval_status | ENUM('Pending', 'Approved', 'Rejected') | Review status | Default Pending |

**Declare of Interest**

The authors declare no competing interest.

# Supplementary Information for:
# PTST: A polar topological structure toolkit and database

Guanshihan Du[1,#], Yuanyuan Yao[2,#], Linming Zhou[1], Yuhui Huang[1,3], Mohit Tanwani[4], He Tian[1,5,6,7], Yu Chen[2], Kaishi Song[2], Juan Li[8], Yunjun Gao[2], Sujit Das[4], Yongjun Wu[1,3,6,9,*], Lu Chen[2,*], Zijian Hong[1,3,5,7,*]

[1] State Key Laboratory of Silicon and Advanced Semiconductor Materials, School of Materials Science and Engineering, Zhejiang University, Hangzhou, Zhejiang 310058, China

[2] College of Computer Science, Zhejiang University, Hangzhou, Zhejiang 310058, China

[3] Zhejiang Key Laboratory of Advanced Solid State Energy Storage Technology and Applications, Taizhou Institute of Zhejiang University, Taizhou, Zhejiang 318000, China

[4] Materials Research Centre, Indian Institute of Science, Bangalore-560012, India

[5] Center of Electron Microscopy, School of Materials Science and Engineering, Zhejiang University, Hangzhou 310027, China

[6] Institute of Fundamental and Transdisciplinary Research, Zhejiang University, Hangzhou 310058, China

[7] Pico Electron Microscopy Center, Hainan University, Haikou 570228, China

[8] College of Materials Science and Engineering, Zhejiang University of Technology, Hangzhou 310014, China.

[9] School of Engineering, Hangzhou City University, Hangzhou, Zhejiang 310015, China

[#] Equal Contributions

[*] Corresponding authors:

Y. W. (yongjunwu@zju.edu.cn); L.C. (luchen@zju.edu.cn); Z. H. (hongzijian100@zju.edu.cn)

**GL-Transformer Framework**

1. Overall Workflow

Figure S1 illustrates the GL-Transformer pipeline for analyzing ferroelectric polarization data, integrating both self-supervised learning (SSL) and subsequent feature extraction. The raw input consists of three-dimensional spatial coordinates (x,y,z) alongside polarization components (px,py,pz). Each dataset is subdivided into multiple "chunks" along the (x,y) plane and sliced along z, creating sub-blocks. Prior to feeding these sub-blocks into the Transformer encoder, a masking procedure is applied. Once SSL training is complete, attention pooling is employed to generate final feature vectors for each sub-block, which are then aggregated for clustering.

2. Masking and Data Augmentatio

The code uses row-wise and span-wise masking to allow the model to learn robust representations. No placeholder token is substituted; instead, masked entries are explicitly set to 0.0 to signal missing information. The steps are:

1) Row-Level Random Masking

- Each row contains [x, y, z, px, py, pz].
- For each row, [x,y,z] may be masked with probability mask_ratio_pos and [px,py,pz] with probability mask_ratio_pol.
- If masked, the corresponding entries are set to 0.0.

2) Span Masking

- With user-defined probabilities (e.g., x_span_prob, y_span_pro, z_span_pro), the code identifies a continuous region along one axis (x, y, or z).
- Any rows whose coordinate falls within that region have their polarization components [px,py,pz] set to 0.0.

3) Random Sign Flipping

- A fraction of the [px,py,pz] values may be multiplied by −1 (with probability flip_prob), encouraging the model to learn orientation-invariant features.

During SSL training, the model attempts to reconstruct the masked entries while also predicting block-level statistics (e.g., mean, min, max of [px,py,pz]), forcing it to capture both local details and global domain characteristics.

3. Transformer Encoder with Positional Encoding

Positional encoding is introduced to preserve spatial context:

1) Coordinate Normalization and Encoding
   - Each [x,y,z] is normalized by dividing by a chosen resolution (e.g., 100).
   - Multiple sinusoidal frequencies are applied to embed spatial coordinates, resulting in a higher-dimensional representation appended to the original row data.

2) Transformer Layers
   - After concatenation of normalized coordinates, polarization data, and positional encoding, a linear projection maps the input to a dimension d_model.
   - This embedding is processed by multiple layers of multi-head attention and feed-forward modules, implemented in *PyTorch* with batch_first=True.

4. Attention Pooling and Feature Extraction

Once self-supervised training (i.e., row reconstruction and block-level statistics) is complete, the model switches to inference mode, where only the Transformer encoder and attention pooling are retained. The pooling mechanism assigns weights to each token embedding and then sums these weighted embeddings into a single vector of size (1×d_model). Each sub-block undergoes this process, and the resulting embeddings can be merged via a secondary aggregator network (another small Transformer) to form one unified vector per sample. These sample-level embeddings are then clustered using agglomerative clustering (Ward linkage).

5. Shared Weight

The same encoder (including positional encoding layers) is used in both the SSL phase and the feature extraction phase. By preserving these weights, the model leverages the knowledge gained from reconstructing masked entries and inferring block-level attributes, ultimately producing richer representations for clustering and further analyses.

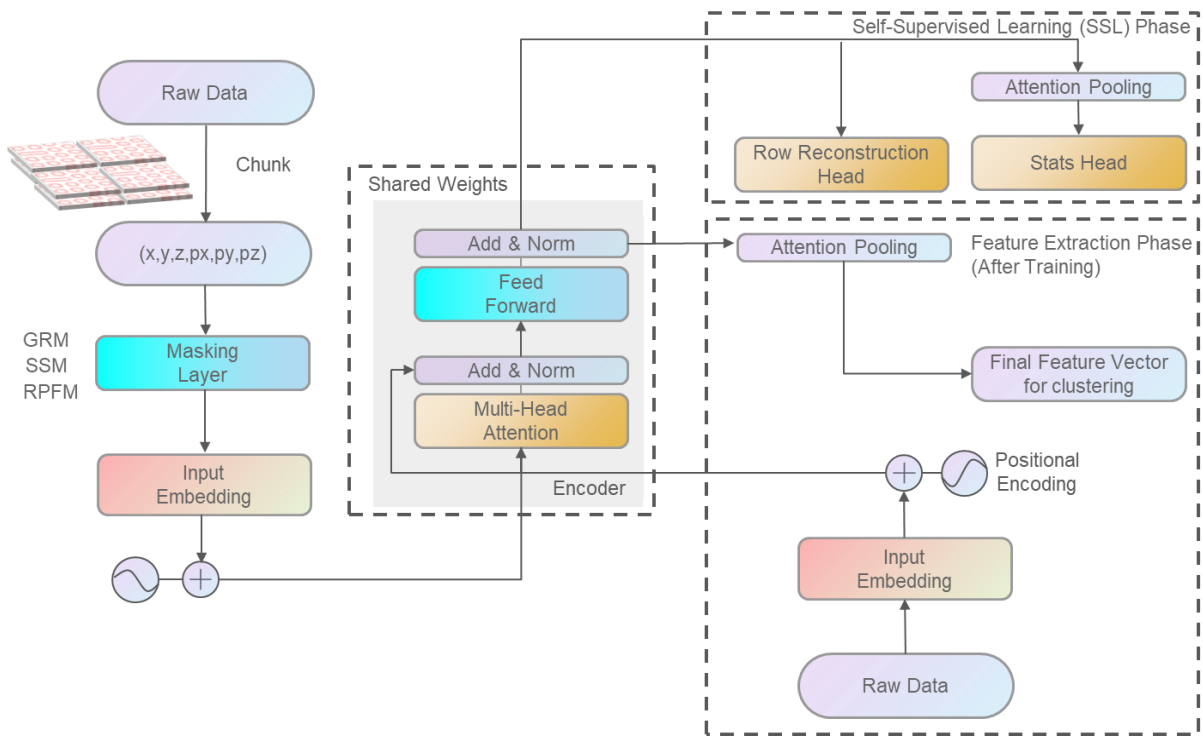

**Figure S1.** Schematic of the GL-Transformer framework used for polarization data analysis. Raw polarization data containing spatial coordinates (x,y,z) and polarization components (px,py,pz) are first chunked and passed through a masking layer before input embedding. The self-supervised learning (SSL) phase involves a transformer encoder with multi-head attention and feed-forward layers, performing row reconstruction and statistical prediction tasks. After SSL training is completed, the feature extraction phase begins, where attention pooling is applied to the encoded features to obtain the final feature vector for clustering. Shared weights are maintained throughout both phases.

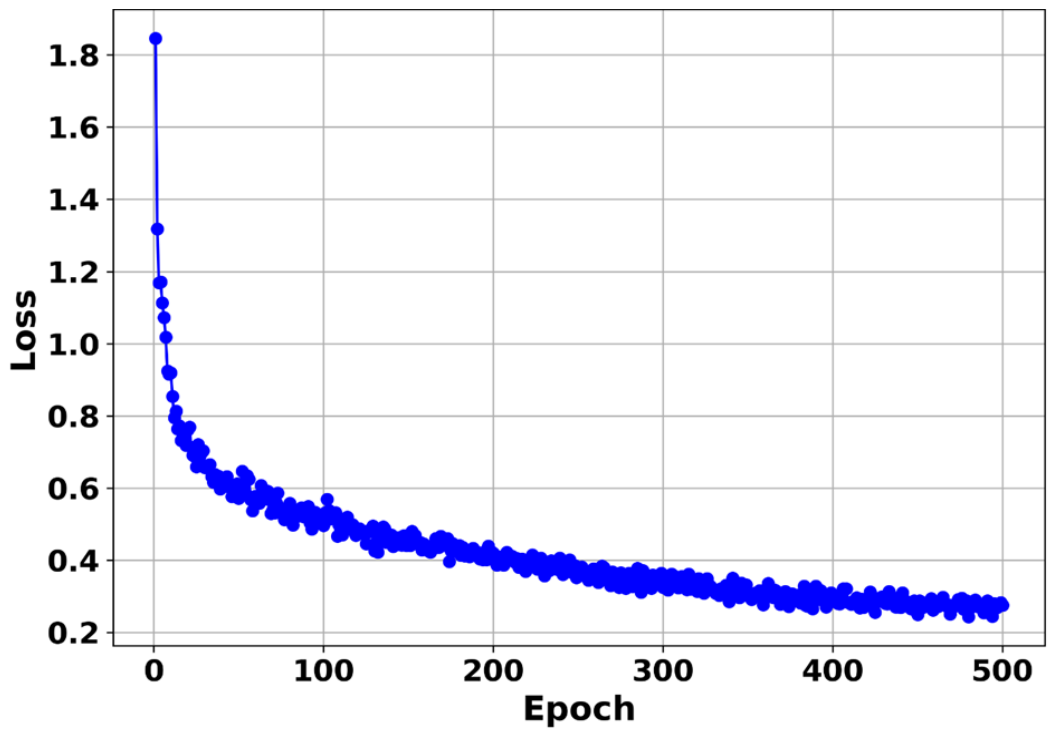

**Figure S2.** Self-supervised training curve of the GL-Transformer model. The vertical axis indicates the loss, while the horizontal axis shows the training epochs.

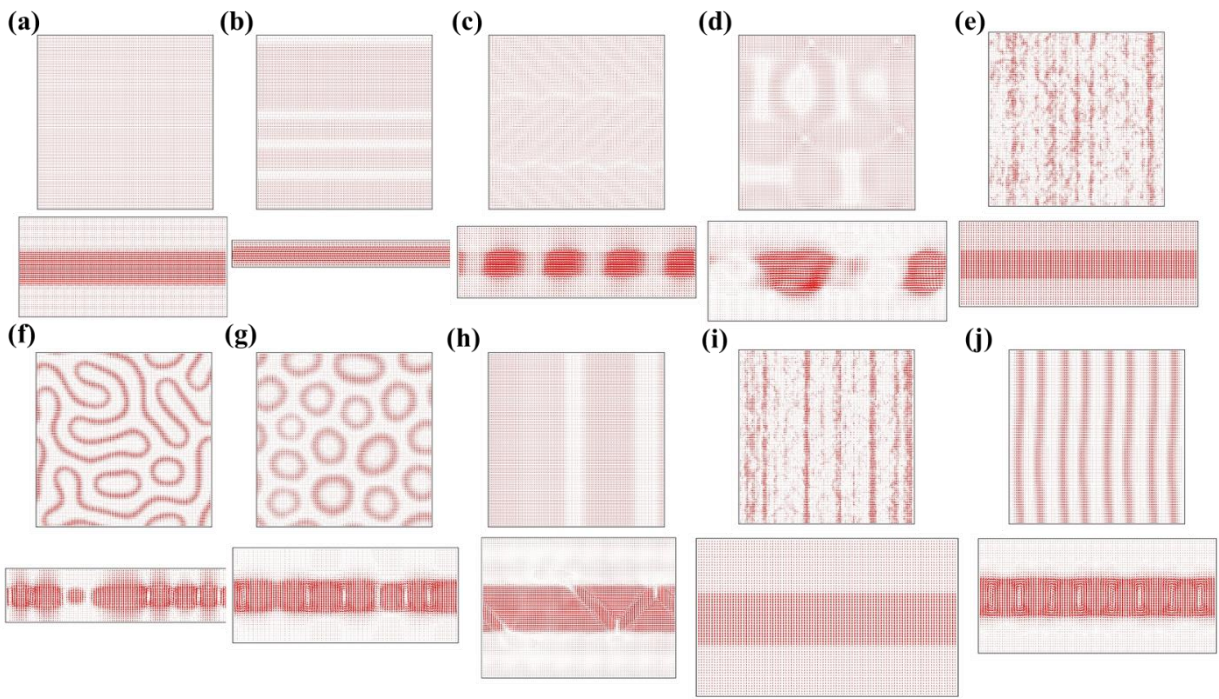

**Figure S3.** The planar view of the in-plane polarization magnitude and the cross-sectional distribution of the out-of-plane polarization component corresponding to the structures shown in Fig. 3(a–j), respectively.

**Table S1. GL-Transformer Model Parameters**

| Component | Parameter Name | Value/Description |
|---|---|---|
| Positional Encoding | max_res | 100 |
| | num_freqs | 4 (Multi-frequency sinusoidal encoding) |
| Transformer Encoder | d_model | 128 |
| | nhead | 8 (Multi-head attention) |
| | num_layers | 6 (Stacked layers) |
| | dim_feedforward | 256 |
| | batch_first | True (batch dimension is first) |
| Self-Supervised Heads | Row Reconstruction | Predicts masked [x,y,z,px,py,pz] set to 0.0 in the input |
| | Statistical Prediction | Estimates 12D block-level statistics (mean, std, min, max for [px,py,pz]) |
| Masking | Row-Level Masking | Randomly sets coordinates or polarization components to 0.0, with user-defined probabilities (mask_ratio_pos, mask_ratio_pol) |
| | Span Masking | Identifies continuous intervals along x, y, or z for additional masking of [px,py,pz] |
| | Random Sign Flip | Multiplies [px,py,pz] by -1.0 with probability flip_prob |
| Attention Pooling | att_pool | Single linear layer for weight scoring, followed by softmax and a weighted sum of token embeddings |
| Aggregator Transformer | num_layers | 2 (Hierarchical merging of sub-block embeddings) |
| | d_model | 128 |
| | nhead | 4 |
| | dim_feedforward | 256 |